\documentclass[fleqn,twoside]{article}
\usepackage{insa}

\usepackage{graphicx}
\usepackage[figuresright]{rotating}

\title{Strong gauge-boson scattering at LHC}

\author{Saurabh D. Rindani \address[PRL]{Theoretical Physics Division \\
	Physical Research Laboratory\\
	Navrangpura, Ahmedabad 380009}%
      } 

\begin{document}
\thispagestyle{empty}

\begin{abstract}
In the standard model with electroweak symmetry breaking through the
Higgs mechanism, electroweak gauge-boson scattering amplitudes are large
if the Higgs boson is heavy, and electroweak gauge interactions become
strong.
In theories with electroweak symmetry
breaking through alternative mechanisms, there could be a strongly
interacting gauge sector, possibly with resonances in an accessible
energy region. In general, the
scattering of longitudinally polarized massive gauge bosons can give information on the mechanism
of spontaneous symmetry breaking. At energies below the symmetry
breaking scale, the equivalence theorem relates the scattering
amplitudes to those of the ``would-be'' Goldstone modes. In the absence of
Higgs bosons, unitarity would be restored by some new physics which can
be studied through $WW$ scattering. Some representatives models
are discussed. Isolating $WW$ scattering at a
hadron collider from other contributions involving $W$ emission from
parton lines needs a good understanding of the backgrounds. 
Resonances, if they exist below about a TeV, would be feasible of
observation at the LHC.
\end{abstract}

\maketitle

% DEFINITIONS

\def\bm {\boldmath}
\def\slq{q \hspace{-1ex}/}
\def\slk{k \hspace{-1ex}/}
\def\bea{\begin{eqnarray}}
\def\eea{\end{eqnarray}}
\def\be{\begin{equation}}
\def\ee{\end{equation}}
\def\beq{\begin{equation}}
\def\eeq{\end{equation}}
\def\dsp{\displaystyle}

\def\gv{g_V}
\def\ga{g_A}
\def\pl{P_L}
\def\plbar{\overline P_L}
\def\pt{P_T}
\def\ptbar{\overline P_T}
\def\peff{P^\textrm{eff}_L}
\def\p{\partial}
\def\sbar{\bar s}

\newcommand{\comment}[1]{}

%
% Definitions
%
\def \smgroup{$SU(2)_L\times U(1) \times SU(3)_C~$}
\def \be{\begin{equation}}
\def \ee{\end{equation}}
\def \ba{\begin{array}}
\def \ea{\end{array}}
\def \bea{\begin{eqnarray}}
\def \eea{\end{eqnarray}}
\def \ol{\overline}
\def \o{\overline}
\def \pa{\partial}
\def \phidot{\stackrel{.}{\phi}}
\def \dag{^\dagger}
\def \phip{\phi'}
\def \pdag{^{\prime\dagger}}
\def \half{\frac{1}{2}}
\def \onebyfour{\frac{1}{4}}
\def \onebyroot2{\frac{1}{\sqrt{2}}}
\def \l({\left(}
\def \r){\right)}
\def \lag{{\cal L}}
\def \ra{\rightarrow}
\def \ba{\begin{array}}
\def \ea{\end{array}}
\def \be{\begin{equation}}
\def \ee{\end{equation}}
\def \bea{\begin{eqnarray}}
\def \eea{\end{eqnarray}}
\def \eebar{$e^+e^-$}
\def \half{\frac{1}{2}}
\def \ra{\rightarrow}
\def \o{\overline}
\def \lag{{\cal L}}
\def \calL{{\cal L}}
\def \lam{\lambda}
\def \L{\Lambda}
\def \f{\frac}
\def \inf{\infty}
\def \p1{\phi_1}
\def \ph2{\phi_2}
\def \ZPC{{\it Zeit. Phys.~}}
\def \PRD{{\it Phys. Rev. D~}}
\def \PRL{{\it Phys. Rev. Lett.~}}
\def \PLB{{\it Phys. Lett. B~}}
\def \NPB{{\it Nucl. Phys. B~}}
\def \JHEP{{\it JHEP~}}

\section{Introduction}

The standard model (SM) of strong, electromagnetic and weak interactions
is based on local gauge invariance under the group $SU(2)_L\times U(1)
\times SU(3)_C$, spontaneously broken to $U(1)_{\rm em}\times SU(3)_C$. 
The spontaneous breaking of gauge invariance gives masses to the gauge
bosons $W^{\pm}$ and $Z$ corresponding to three of the generators of the
gauge group in a way which maintains the renormalizability of the
theory. The gauge bosons corresponding to the $U(1)_{\rm em}$ gauge group
of electromagnetism and the $SU(3)_C$ gauge group of colour remain
massless. The spontaneous braking is also required to give masses to the
fermions in the theory.

Spontaneous symmetry breaking (SSB) in the SM is achieved by means of the
Higgs mechanism. An $SU(2)_L$ doublet of spin-0 fields is present in the
theory, and the configuration which minimizes the action corresponds to
a nonzero value for the neutral component of the doublet. Thus, the
ground state (vacuum state) 
of the theory is not invariant under the $SU(2)_L$ group,
and the gauge symmetry is broken spontaneously. Since it is the neutral
component of the scalar doublet which has a nonzero vacuum expectation
value, $U(1)_{\rm em}$ remains an unbroken symmetry.

As a result of SSB, three of the four degrees of freedom of
the complex scalar doublet, get absorbed resulting in the longitudinal
degrees of freedom of the (massive) gauge bosons. These are the
so-called ``would-be Goldstone'' modes, since the Goldstone mechanism
for a spontaneously broken {\it global} invariance would have required
these modes to propagate as massless particles. There remains one scalar
field, the particle corresponding to which becomes massive.
This massive spin-0 particle is known as the Higgs boson.

The introduction of spin-0 fields in the theory, while successful in
breaking the electroweak gauge invariance to generate masses for gauge
bosons and fermions, leads to certain problems. The main problem is that
the mass of the Higgs boson cannot be made small in a natural way. 
As a result, alternative proposals for have been considered for the
mechanism of spontaneous symmetry breaking, some of which do not result
in a light Higgs boson. The issue of the mechanism of spontaneous
electroweak symmetry breaking is thus an important one. If the Higgs
boson is discovered with properties predicted by SM, it would be a proof
of Higgs mechanism. If the Higgs boson is not found, which would imply
that it is either not present or very heavy, alternative mechanisms of
SSB would have to be taken seriously.

The Large Hadron Collider (LHC) at CERN, Geneva, which in the very near
future will collide protons on protons at a centre-of-mass energy of 14
TeV, is expected to unravel the
mystery of the mechanism of SSB of the
local gauge invariance responsible for the electroweak interactions.
If the SM Higgs boson exists, and is light, it will be
found. Alternatively, if the symmetry breaking is not due to a
straightforward Higgs mechanism, or if the Higgs boson is not light, LHC 
should show other effects, particularly in scattering of weak gauge
bosons. Scattering of longitudinally polarized weak bosons play a
dominant role, and in the absence of a light Higgs boson, it is expected
to become strong. Thus $W_LW_L$, $W_LZ_L$ and $Z_LZ_L$
scattering (where the subscript $L$
denotes the longitudinal polarization mode), will be an important tool to
study the mechanism of SSB.

Here we discuss the importance of studying gauge-boson scattering as a
means of gaining knowledge about the electroweak symmetry breaking  mechanisms,
and also the feasibility of carrying out such a study at LHC. 
There have been a number of studies on strong gauge-boson scattering
over the years. Since it is next to impossible to do an exhaustive review
here, we will be forced to be selective.

We begin by discussing general features of scattering of massive spin-1
particles and specialize to the case when these particles are the gauge
bosons of a spontaneously broken local gauge theory. 

\subsection{Longitudinal gauge-boson scattering}

Amplitudes for scattering in theories with massive spin-1 particles can 
have bad high-energy behaviour. The propagator for a massive spin-1 field has a
term which does not decrease with increasing momentum, leading to bad
high-energy behaviour in amplitudes involving exchange of these vector
particles.
This can make the theory non-renormalizable.
In theories with SSB, using the Higgs
mechanism, renormalizability is
maintained because 
Higgs exchange can cancel the bad high-energy behaviour arising from
vector exchange. Amplitudes with massive vector particles in the initial
and final states have bad high-energy 
behaviour because the polarization vector has the form
\beq
\epsilon_L^\mu (k) \approx \frac{k^\mu}{m_V}.
\eeq
In the unitarity gauge, scattering of longitudinally polarized
gauge bosons would have the largest amplitudes at high energy.
In the high-energy limit, the amplitude for $W^+_LW^-_L$ takes the form
\be\label{gaugeamp}
{\cal M}^{\rm gauge} = -\frac{g^2}{4\, m_W^2}\;u + {\cal
O}((E/m_W)^0),
\ee
where $u$ is the Mandelstam variable.
Again, if the mass of the vector fields arises through a Higgs
mechanism, the bad high-energy behaviour is cured. 
The contribution of $s-$ and $t-$ channel Higgs exchanges is
\be 
\begin{array}{lcl}
{\cal M}^{\rm Higgs}& =& -\frac{g^2}{4\, m_W^2}\;\left[
\frac{(s-m_W^2)^2}{s-m_H^2} + 
 \frac{(t-m_W^2)^2}{t-m_H^2}\right]\\
&\approx & \frac{g^2}{4\,m_W^2}\; u,
\end{array}
\ee
in the limit $s>> m_H^2,m_W^2$, which cancels the bad high-energy
behaviour of (\ref{gaugeamp}).
However, if the Higgs mass is too high ($s<<m_H^2$), the amplitude can be 
large, leading to a 
strongly-interacting gauge sector.
The study of $WW$ scattering can thus be important in studying the
mechanism of symmetry breaking. (Here, as well in what follows, we use
$WW$ to denote $WW$, $WZ$ as well as $ZZ$ combinations). 

\subsection{The equivalence theorem}

Vector-boson scattering amplitudes at high energies may be calculated
using the equivalence theorem \cite{Cornwall}.
The equivalence theorem relates the $W_LW_L$ scattering amplitude in the
Feynman-'tHooft gauge to the
amplitude for scattering of the corresponding ``would-be'' Goldstone
scalars at high energy ($s >> m_W^2$). 
Thus,
\be
\begin{array}{l}
M(W_LW_L \to W_LW_L) = M(ww \to ww)
\\
M(Z_LZ_L \to Z_LZ_L) = M(zz \to zz)
\\
M(W_LZ_L \to W_LZ_L) = M(wz \to wz)
\\
M(W_LW_L \to Z_LZ_L) = M(ww \to zz),
\end{array}
\ee
where $w$ and $z$ are respectively the scalar modes which provide the
longitudinal modes for $W$ and $Z$ respectively.
 
Goldstone boson interactions are governed by low-energy theorems
for energy below the symmetry breaking scale ($s << m_{\rm SB}^2$).
Low-energy theorems are analogous to those obtained for $\pi\pi$
scattering in the chiral Lagrangian
\cite{Weinberg}, and depend only on the symmetry of the theory.
Thus, when there is no light Higgs (with $m_H < 1$~TeV or so), 
the low-energy theorems combined with
equivalence theorem can predict $W_LW_L$ scattering amplitudes
from the symmetries of the theory to leading order in $s/m_{\rm SB}^2$.
The specific theory for symmetry breaking then shows up at the next
higher order in $s/m_{\rm SB}^2$.

In the standard model (SM), we have the  relations 
\beq
m_H^2 = -2\mu^2 = 2\lam v^2 = \lam \sqrt{2}/G_F,
\eeq
where $\lam$ is the quartic scalar coupling, and $\mu$ is the mass
parameter in the scalar potential. $v$ is the vacuum expectation value
of the neutral Higgs.
Since $v$ and $G_F$ are fixed from experiment, large $m_H$ means large
$\lambda$. 
 For $\lambda$ \raisebox{-4pt}{$\stackrel{>}{\sim}$} $4\pi$, i.e., 
$m_H$ \raisebox{-4pt}{$\stackrel{>}{\sim}$}
$(4\pi\sqrt{2}/G_F)^{\frac{1}{2}}$,
perturbation theory is not valid. 
 This corresponds to $m_H\approx 1.2$
TeV.

\subsection{Limit from unitarity}

   A limit may be obtained from unitarity, if the tree
amplitudes are to be valid at high energies.
In the absence of the Higgs, the amplitudes violate unitarity at high
energies. The limit is of the order of the symmetry breaking scale.
The Higgs cannot be very heavy, or else the
unitarity limit would be crossed too soon. To see this, write the
old-fashioned scattering amplitude $f_{cm}$, so that $\vert f_{cm}
\vert^2 = \f{d\sigma}{d\Omega}$. Then we have 
\beq
f_{cm} = \f{1}{8\pi\sqrt{s}}{\cal M}.
\eeq
The scattering amplitude has the partial wave expansion
\beq
f(\theta) = \f{1}{k} \sum_l (2l+1)P_l (\cos\theta) a_l,
\eeq
where $a_l = e^{i\delta_l} \sin\delta_l$
is the partial wave amplitude written in terms of the phase shift
$\delta_l$. Unitarity is expressed by the optical theorem relation
\beq
\sigma = (4\pi/k)\,{
\rm Im} f(0).
\eeq
For elastic scattering, the phase shift $\delta_l$ is real, but has a
positive imaginary part if there is inelasticity. A convenient way to
express elastic unitarity is 
\beq
{
\rm Im} [a_l^{-1}] = -1.
\eeq
From this it follows that
\beq 
\vert a_l \vert \leq 1; \;\;\vert {\rm Re}\, a_l\vert \leq \half.
\eeq
The $l=0$ partial-wave unitarity for large $s$ for $WW$ scattering then
gives, in the absence of the Higgs boson,
\beq 
{\rm Re~} a_0\equiv{\rm Re~} \frac{G_Fs}{16\pi\sqrt{2}} < 1/2.
\eeq
Thus, for large enough $s$, unitarity is violated. The best bound comes
from the isospin zero channel $\sqrt{1/6}\, (2W_L^+W_L^- + Z_LZ_L)$, viz.,
\beq
s< \frac{4\pi\sqrt{2}}{G_F}\approx (1.2\; {\rm TeV})^2.
\eeq

In the case of the standard Higgs mechanism, unitarity is restored by the
exchange of a light Higgs boson. For unitarity to hold in the presence
of the Higgs boson, the mass of the Higgs boson should satisfy 
\beq
m_H^2 < \f{4\sqrt{2}\pi}{G_F}.
\eeq
This gives a limit of $m_H < 1.2 $ TeV. Similar limits may be obtained
by considering other partial-wave channels.

If there is no light Higgs, the perturbative unitarity limit is
violated, and the gauge interactions are no longer weak. We then have a
strong gauge sector. This could lead to the possibility of
strong-interaction resonances, as seen in the case of the usual hadrons.

We will know discuss scenarios beyond the standard Higgs mechanism,
where either there is no light Higgs, or the Higgs sector is more
complicated than that in SM, so that the the $WW$ scattering can become
strong.

\section{\bm $W_L W_L$ scattering beyond Higgs mechanism}

As we have seen, absence of a light Higgs boson leads to violation of
unitarity. 
Violation of unitarity may be prevented, or postponed to higher energy,
 in different ways, depending on the
model.
Models can have extra fermions and extra gauge interactions, which give
additional contributions to $W_LW_L$ scattering, e.g., resonances.
An important consideration for many of the theories with a modified
electroweak symmetry-breaking mechanism is conformity with precision
tests. Specifically, they can have an unacceptably large $S$ parameter
of the oblique parameters $S$, $T$, $U$ proposed by Peskin and Takeuchi
\cite{stu}, and a negative contribution to $S$ is needed to restore
agreement with experiment.

It is necessary to discriminate among mechanisms which give new effects
in $WW$ scattering. In particular, their treatment would be different
depending on whether the gauge interaction remains perturbative or not.
One possibility is that of a heavy Higgs,
$m_H>>m_W^2$. In such a case, $WW$ scattering would become large until
the Higgs pole is crossed. If the Higgs mass is above the unitarity
limit, the gauge-boson interaction becomes non-perturbative. This needs
special non-perturbative treatment of gauge-boson interactions, and it
is possible that the strong interaction produces resonances which would
be seen. The second possibility is that there is no Higgs boson, and the
symmetry is broken dynamically. In such a case there is a new sector
with strong interactions, and this could produce its own resonances.
A third possibility is that there is a light Higgs in an extension of
the standard model. Then again two cases arise: One where there are
other Higgs bosons, and the dynamics still remains perturbative for
$WW$ centre-of-mass energy beyond the light Higgs resonance until the
other Higgs resonances are crossed, or could become
strong if the remaining Higgs bosons are heavy. These possibilities are
discussed in the context of 
two-Higgs doublet models in \cite{cheung,randall}. 
The other case is when the light Higgs is
light because it is a pseudo-Goldstone boson of a higher global
symmetry, broken by the SM interactions \cite{giudice}. 
In either case, $WW$ scattering
would be interesting to study in spite of a light Higgs boson.

We discuss a few of these models below. However, we begin with a
model-independent approach.

\subsection{Electroweak Chiral Lagrangian}

In view of the variety of models, it would be preferable to have a
unified, model-independent description of $W_LW_L$ scattering. Effective
Lagrangians may be used to provide such a description, valid up to a
certain cut-off $\Lambda$.
A no-resonance scenario is described in an electroweak chiral Lagrangian (EWCL)
model, where one writes effective bosonic operators \cite{EWCL} in a
series of increasing dimensions.  The coefficients will
have inverse powers of $\Lambda$.
The effective Lagrangian is a low-energy expansion, and at the lowest
non-trivial order, corresponding to operators of the lowest dimension,
the scattering amplitudes for longitudinal gauge bosons, on using the
equivalence theorem, would satisfy the low-energy theorems. However,
coefficients of the higher-dimensional operators have to be fixed either
on the basis of some underlying theory, or, in a purely phenomenological
approach, from experiment.

Since the chiral Lagrangian is valid at low energies, an extrapolation
to higher energies would lead to violation of unitarity.
Unitarization can be built in in a somewhat {\it ad hoc} fashion 
by the use of Pad\' e approximants, also known as the Inverse Amplitude
Method (IAM), or
the K-matrix method and this can generate a resonant behaviour \cite{DH}.
The $N/D$ method, developed for strong interactions \cite{chew}, has also been
applied for unitarization of electroweak amplitudes
\cite{oller}.

 Terms in the chiral Lagrangian must respect  
$SU(2)_{L} \times U(1)$ gauge symmetry, which is broken spontaneously. 
Experimental constraints require that the Higgs
sector also approximately respect a larger $SU(2)_L \times SU(2)_C$
symmetry, where the second factor $SU(2)_C$ corresponds to the so-called
custodial symmetry. The custodial symmetry is broken by the
Yukawa couplings and the $U(1)$ gauge couplings.
The global symmetry could in principle be even larger.

The chiral Lagrangian is thus constructed using the dimensionless
unitary
unimodular matrix field $U(x)$, which transforms under $SU(2)_L \times
SU(2)_C$ as $(2,2)$. This corresponds to a non-linear realization of the
symmetry. $U(x)$ may be represented in terms of real (would-be
Goldstone) fields $\pi_a$, ($a=1,2,3$), as follows:
\be
U(x) = \exp{\frac{i}{v}\sum_{a=1}^3 \pi_a \tau_a },
\ee
where $\tau_a$, $a=1,2,3$, are the usual $2\times 2$ Pauli
matrices, and $v$ is the symmetry breaking scale\footnote{These $\pi_a$ should not be confused with the
$\pi$-meson fields. The $w$, $z$ scalar fields, corresponding to the
longitudinal polarization modes of the gauge bosons $W$, $Z$, are linear
combinations  of the $\pi_a$.}. 

The covariant derivative of $U(x)$ is
\begin{equation}
D_{\mu}U = \partial_{\mu} U + i g \frac{\vec{\tau}}{2} \cdot
\vec{W_{\mu}}
U -i g' U \frac{\tau_3}{2} B_{\mu}.
\end{equation}

Further, the basic building blocks which are $SU(2)_{L}$ covariant
and $U(1)_{Y}$ invariant are 
\begin{equation}
T  \equiv  U \tau_{3} U^{\dag}\ ,\; \; \; \;
V_{\mu}  \equiv  (D_{\mu} U ) U^{\dag} ,
\ee
\be
W_{\mu\nu}  \equiv  \partial_{\mu} W_{\nu} - \partial_{\nu} W_{\mu} +
ig[W_{\mu}, W_{\nu}]
\end{equation}
where $T$\,,\,$V_{\mu}$ and $W_{\mu\nu}$ have dimensions zero, one, and
two
respectively.

 Terms in the chiral Lagrangian in the
$M_{H} \rightarrow \infty$ limit of the linear
theory at tree level are:
\bea
{\cal L}_{0}& = & \frac{1}{4} v^2 Tr[(D_{\mu}U)^{\dag}(D^{\mu}U)]
\nonumber \\
&&- \frac{1}{4} B_{\mu\nu}B^{\mu\nu} - \frac{1}{2} TrW_{\mu\nu}W^{\mu\nu},
\eea
where $v$ is the symmetry breaking scale 
and $B_{\mu\nu} \equiv \partial_{\mu} B_{\nu} - \partial_{\nu} B_{\mu}$.
 The first term has dimension two, while
the second two (kinetic energy) terms have dimension four.
 The gauge  couplings to the quarks and leptons must also be added.
 The Yukawa couplings of the quarks and leptons should also be included
in the symmetry breaking sector.

An additional dimension-two operator is allowed by the
$SU(2)_{L} \times U(1)$ symmetry:
\be
{\cal L}_{1}^{\ '} =  \frac{1}{4} {\beta}_1 g^{2} v^{2}
[Tr(TV_{\mu})]^{2}.
\ee
This term does not emerge from the $M_{H} \rightarrow \infty$
limit of the renormalizable theory at tree level. It violates the
$SU(2)_{C}$
custodial symmetry even in the absence of the
gauge couplings. 
It is the low-energy description of 
custodial-symmetry breaking physics which has been integrated out,
at
energies above roughly $\Lambda \equiv 4{\pi}v \simeq 3$~TeV. 
 At tree level, ${\cal L}_{1}^{\ '}$
contributes to the deviation of the $\rho$ parameter from
unity.

 At the dimension-four level, there are a variety of new operators that
can be written down. 
Making use of the equations of motion, and 
 restricting attention
to CP-invariant operators, the list can be reduced to eleven independent
terms \cite{EWCL}:

\begin{equation}
\begin{array}{l}
{\cal L}_{1}  =  \frac{1}{2}\alpha_1 gg' B_{\mu\nu}
Tr(TW^{\mu\nu})\\
{\cal L}_{2}  = \frac{1}{2} i \alpha_2 g' B_{\mu\nu} Tr(T[V^{\mu},
 V^{\nu}])   \\
{\cal L}_{3}  =  i \alpha_3 g  Tr(W_{\mu\nu}[V^{\mu}, V^{\nu}])
\\
{\cal L}_{4}  =  \alpha_4 [Tr(V_{\mu}V_{\nu})]^2   \\
{\cal L}_{5}  =  \alpha_5 [Tr(V_{\mu} V^{\mu})]^2 \\
{\cal L}_{6}  =  \alpha_6 Tr(V_{\mu} V_{\nu})Tr(TV^{\mu})
Tr(TV^{\nu})  \\
{\cal L}_{7}  =  \alpha_7 \  Tr(V_{\mu} V^{\mu})Tr(TV_{\nu})
Tr(TV^{\nu}) \\
{\cal L}_{8}  =  \frac{1}{4} \alpha_8 \  g^2 \ [Tr(TW_{\mu\nu})]^2
 \\
{\cal L}_{9}  =  \frac{1}{2} i \alpha_9 g
Tr(TW_{\mu\nu})Tr(T[V^{\mu},
V^{\nu}]) \\
{\cal L}_{10}  = \frac{1}{2} \alpha_{10}
[Tr(TV_{\mu})Tr(TV_{\nu})]^2
  \\
{\cal L}_{11}  =  \alpha_{11} g {\epsilon}^{\ \mu\nu\rho\lambda}\
Tr(TV_{\mu})Tr(V_{\nu}W_{\rho\lambda}) 
\end{array}
\end{equation}

The EWCL incorporates the low-energy theorems, and therefore the
piece ${\cal L}_0$ is completely fixed in terms of the symmetry-breaking
scale $v$. However, the coefficients of the higher-dimensional operators
have to be determined from experiment. Alternatively, they can be
matched with a suitable underlying theory of one's choice.

The pieces of the EWCL contributing to gauge-boson two-point functions
have been tested in precision tests, and these contribute to the
oblique parameters, $S$, $T$ and $U$ \cite{stu}.
 ${\cal L}_{1}^{'}$, ${\cal L}_{1}$ and ${\cal L}_{8}$ are
directly related
to the $S$, $T$ and $U$ parameters.
By setting the Goldstone boson fields to zero in these
operators, which is equivalent to going to the unitarity gauge, we get
\be\label{s}
S  \equiv  - 16 \pi \frac{d}{dq^2}{\Pi}_{3B}(q^2){|}_{q^2=0}  =  -16
\pi
{\alpha}_1 ,
\ee
\be\label{t}
\alpha T  \equiv  \frac{e^2}{c^2s^2m_{Z}^2} ({\Pi}_{11}(0) -
{\Pi}_{33}(0))
 =  2 g^2 \beta_1,
\ee
\be
U  \equiv  16 \pi \frac{d}{dq^2}[{\Pi}_{11}(q^2) - {\Pi}_{33}(q^2)]
{|}_{q^2=0}  =  -16 \pi {\alpha}_8.
\ee
The $\Delta\rho( \equiv \rho -1)$ parameter is related to $T$ by
 $\Delta\rho_{new} = \Delta\rho
- \Delta\rho_{SM} = \alpha T$, where $\Delta\rho_{SM}$ is the
contribution
arising from standard model corrections.

These relations have been  used  in \cite{bagger} to study the possible
values of the EWCL coefficients in the presence of a SM Higgs boson with
a mass larger than the electroweak precision measurement limits. They
find the following 68\% CL allowed ranges for $S$ and $T$, which
could be converted to the ranges for the parameters $\alpha_1, \beta_1$ 
using the eqns.
(\ref{s}) and (\ref{t}) above:
\be
\begin{array}{cc}
-0.37 < S < -0.17, & 0.34 < T < 0.58
\end{array}
\ee
The smallness of the parameter $T$ can be understood in terms of the
approximate $SU(2)_C$ custodial symmetry. This approximate symmetry 
implies that the coefficients $\alpha_{2,6,7,8,9,10,11}$ will be
sub-dominant relative to the custodial-symmetry preserving ones.
Gauge-boson scattering can be considered to be dominated by the two
coefficients $\alpha_4$ and $\alpha_5$.

\begin{figure}[htb]
\centering
\includegraphics[width=5.5cm]{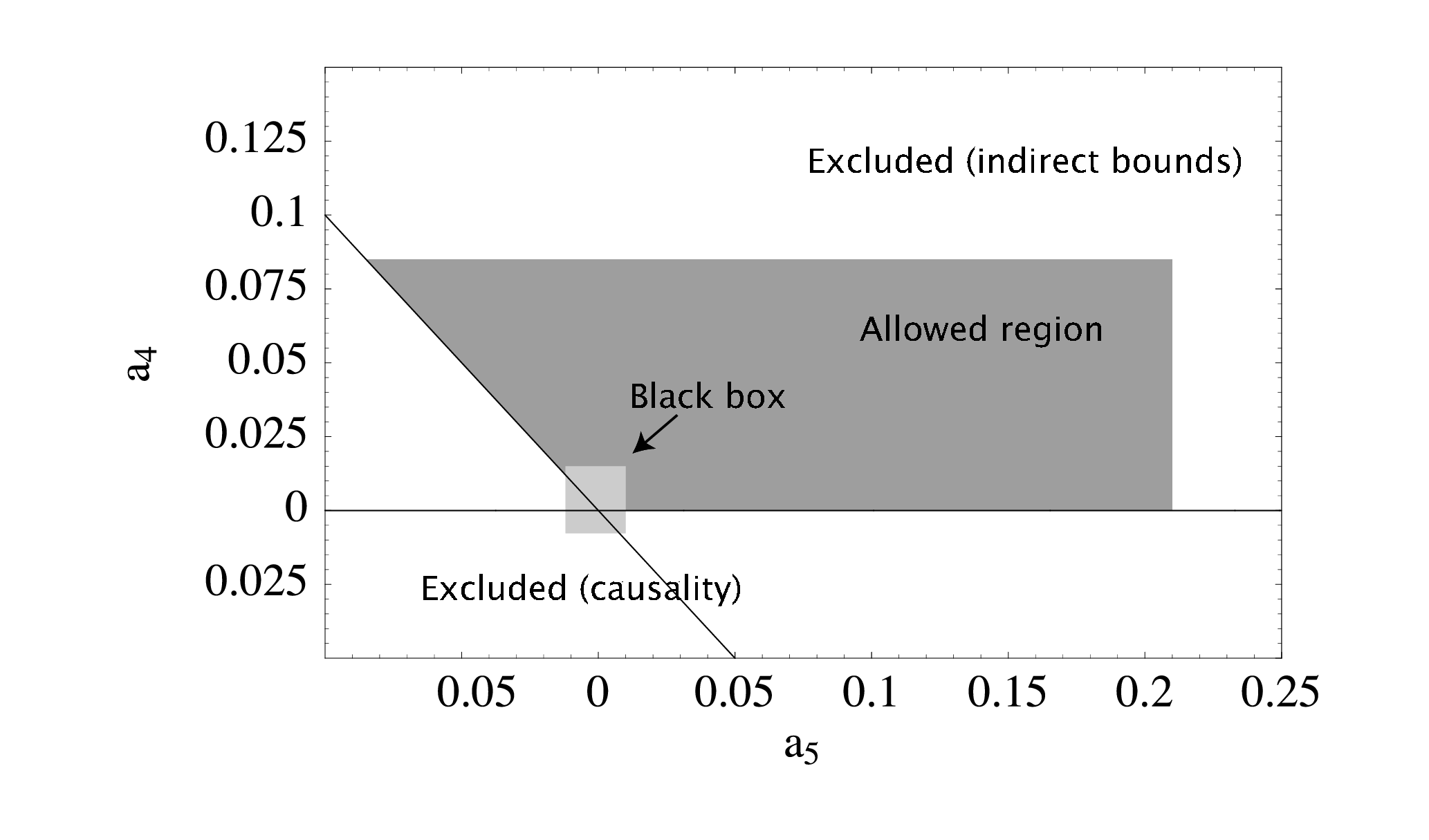}
\caption{The region of allowed values in the $\alpha_4-\alpha_5$ plane
($a_4-a_5$ plane, in the notation of \cite{fabb}), combining indirect
bounds and causality constraints, is shown in gray. Also depicted is the
region below which LHC will not be able to resolve the coefficients
(``black box''). (From \cite{fabb}).}
\label{causality}
\end{figure}

There are constraints from unitarity, analyticity and causality of
scattering amplitudes on the effective
Lagrangian parameters \cite{distler,vecchi}. The requirement of
unitarity of the theory forces the cutoff of the EWCL to be about
$\Lambda<1.2$~TeV, but does not impose any constraints on the
coefficients $\alpha_i,\beta_1$. Causal and analytical structure of the
amplitudes leads to bounds on the possible values of $\alpha_4$ and
$\alpha_5$. These were first obtained in the context of chiral Lagrangians for
strong interactions \cite{pham}, and have been extended to the
electroweak case. The second derivative with respect to the
centre-of-mass energy of the forward elastic scattering amplitude of two
Goldstone bosons is bounded from below by a positive integral of the
total cross section for the transition $2\pi\to {\rm all}$ \cite{distler}.
However, it was shown by Vecchi \cite{vecchi} that stronger bounds are
obtained from causality (we refer the reader to \cite{vecchi} for
details):
\be 
\alpha_4(\mu) \geq \frac{1}{12\,(4\pi)^2}\,\ln\frac{\Lambda^2}{\mu^2},
\ee
\be
\alpha_4(\mu) + \alpha_5(\mu) \geq
\frac{1}{8\,(4\pi)^2}\,\ln\frac{\Lambda^2}{\mu^2}. 
\ee
Fig. \ref{causality}, taken from \cite{fabb} shows the allowed regions
in the plane of $\alpha_4$ and $\alpha_5$ after including constraints
from causality and indirect bounds from precision experiments. The
figure also includes a ``black box'', the region inaccessible at LHC
(see below).

Further constraints on the parameters can be obtained from triple and
quartic gauge boson couplings through direct experimental measurement.
The quartic couplings, which 
correspond to gauge-boson scattering, were studied in the LHC context in
\cite{eboli1,eboli2}, where the second reference cited considered
six-fermion final states at orders $\alpha_{\rm em}^6$ and $\alpha_{\rm
em}^4\alpha_s^2$. It was found in \cite{eboli2} that with an integrated
luminosity of 100 fb$^{-1}$, LHC could place the following 99\% CL
bounds on the EWCL parameters $ \alpha_4,\alpha_5$:
\be\label{LHCsens}
-7.7 \times 10^{-3} < \alpha_4 < 15 \times 10^{-3},\\
\ee 
\be
-12 \times 10^{-3} < \alpha_5 < 10 \times 10^{-3}.
\ee 
These bounds correspond to assuming one parameter nonzero at a time.

In EWCL, one can build in unitarity at a given order by using a
non-perturbative modification, which reduces to the original amplitude
at the perturbative level \cite{DH}.
One writes a low-energy expansion as $a(s) = a^{\rm LET}(s) +
a^{(1)}(s)$, where $a^{\rm LET}$ represents the amplitude given by the
low-energy theorems.
At the lowest order the Pad\' e approximant gives 
\be
a^{\rm Pade} (s) = \frac{a^{\rm LET}(s)}{1-\frac{a^{(1)}(s)}{a^{\rm
LET}(s)}}.
\ee
The $K$ matrix method, which is simpler, consists in multiplying the
amplitude by $-i$ and adding it to the denominator, so that the
condition Im~$a^{-1}$ is satisfied. It gives
\be
a^K (s) = \frac{a^{\rm LET}(s) + {\rm Re}a^{(1)}(s)}{1-i( a^{\rm LET}(s)
+ {\rm Re}a^{(1)})}.
\ee
Both satisfy unitarity by construction to the relevant order.

A recent suggestion \cite{kilian} has been to add resonances to the
EWCL, in appropriate isospin channels so as to preserve custodial
$SU(2)$ symmetry, parametrizing them simply by their masses and widths.
Ref. \cite{kilian} uses a simple $K$ matrix unitarization. With an
appropriate off-shell extension of the interactions, the authors have
implemented the process of $WW$ scattering in an event generator for six
final-state fermions, {it viz.}, {\tt WHIZARD} \cite{whizard}. 

\subsection{Technicolor models}

Technicolor models are of the earliest proposals for dynamical breaking of
the electroweak symmetry \cite{techni}, without the introduction of
elementary scalar fields. In analogy with QCD, local $SU(N)$
gauge invariance among technifermions in the fundamental representation
leads to condensates which act as effective scalar fields and are
responsible for breaking of the electroweak symmetry. The relevant
coupling in technicolor theories is scaled up as compared to QCD to
account for the electroweak scale. Thus, scattering of longitudinal
gauge bosons can be strong. 

The basic problem with technicolor, even
when ordinary fermions acquire mass via an extended technicolor (ETC) sector
\cite{etc}, is that of large flavour-changing neutral currents and a
large contribution to the $S$ parameter. A proposal to address this
problems has been walking technicolor (WT) \cite{wt} (in contrast to the
usual theories which have renormalization group ``running''). Walking dynamics
helps suppressing flavour-changing neutral currents without preventing
ETC from yielding realistic fermion masses. Certain WT models are also
in agreement with constraints imposed by precision electroweak data
\cite{foadi}, because of the fact that walking dynamics itself can lower
the contribution to the $S$ parameter compared to a running theory.
Various other modifications of WT theories have been considered recently
to address different problems, and it seems that it is possible to
construct working models of WT.

The scattering of Goldstone bosons representing the longitudinal gauge
bosons was examined recently in \cite{foadi2,fabb}. In \cite{foadi2},
the authors examine to what extent the violation of unitarity in $WW$
scattering can be delayed in the case of a realistic WT model which
satisfies existing phenomenological constraints. They consider three
scenarios with values of the oblique parameter $S$ 0.1, 0.2 and 0.3.
Making use of Weinberg sum rules relating masses ($M_V$, $M_A$) and 
decay constants ($F_V$, $F_A$) of
vector and axial-vector resonances, they find that unitarity violation
can be delayed to as much as 3.7 TeV for $M_A=1.5$~TeV.
However, there is a critical value of the coupling $g\equiv 2M_V^2/F_V^2$ 
above which the theory violates unitarity at a much lower energy. Thus
in the latter case, a spin-0 isospin-0 resonance would be required. On
the other hand, for values of $g$ below the critical value, the theory
may very well be Higgsless. However, it is entirely possible that the
resonance being heavy may be difficult to observe. 

%\subsection{Little Higgs models}
\subsection{Beyond SM scenarios with a light Higgs}

There are two categories of theories which can have a light Higgs boson, with
new physics in $WW$ scattering above the light Higgs resonance. The
first category corresponds to multi-Higgs models, where more than one
doublet of scalar fields is introduced. In such a case $WW$ scattering
amplitude would grow after crossing the light Higgs resonance, but it
could remain weak, if the heavier Higgs lies below the unitarity limit
\cite{cheung,randall}. 
The other category corresponds to a scenario where there is one light
Higgs boson, which is light because it is a pseudo-Goldstone boson
corresponding to the breaking of a global symmetry. The Higgs could get
mass only at the loop level.  
Above the light Higgs resonance, the $WW$
scattering amplitude could be strong, and the resulting theory could
have other resonances.
The Little Higgs models, together with theories with holographic Higgs,
are studied in \cite{giudice} under the category of ``strongly-interacting
light Higgs'' theories. 

In such theories, the new sector responsible for electroweak symmetry
breaking, characterized by two parameters, a coupling $g_\rho$ and a
scale $m_\rho$ describing the mass of heavy physical states. The
coupling $g_\rho$, while larger than a typical SM coupling, is
restricted to be below $4\pi$ to ensure that loop expansion remains
dependable. 

Experimental tests of such models would be observation of new states in
addition to the light Higgs boson. However, even in the absence of new
states, $W_LW_L$ scattering amplitudes would be found to grow as
$E^2/f^2$, where $f\equiv m_\rho/g_\rho$, since the
Higgs boson moderates the high-energy behaviour only partially. Thus a
study of gauge-boson scattering would prove useful. It is also
predicted that the there would be strong double-Higgs production through
$W_LW_L\to HH$.

\subsection{Higgsless models}

A number of Higgsless models have been proposed recently
\cite{Csaki,Nomura}.
In these models, 
symmetry breaking is achieved by appropriate boundary conditions.
The models  differ in spatial dimensions, 5 in the original versions, 4 in the
``deconstructed'' versions.
They also differ in embedding of SM fermions.

A feature common to such theories in five dimensions is that new 
weakly coupled particles, {\it viz.}, the Kaluza-Klein excitations of
the gauge bosons appear at TeV scale and cancel the bad high-energy
behaviour of $WW$ scattering amplitudes, and thus postpone unitarity
violation.
A version of the model with modified fermion sector can raise the scale of 
unitarity violation by at least a
factor of 10 without running into conflict with precision electroweak
constraints.

In the absence of Higgs, new massive vector boson (MVB) propagators contribute
to $WW$ scattering.
The bad high energy behaviour of $WZ$
scattering, for example, is cancelled by the contribution of the MVBs
because of coupling constant sum rules \cite{higgsless}:
\bea
&g_{WWZZ} =  g^2_{WWZ} + \sum_i (g^{(i)}_{WZV})^2,&\\
&2(g_{WWZZ} - g^2_{WWZ})(M_W^2 + M_Z^2 ) + g^2_{WWZ}\frac{M_Z^4}{M_W^2}
&\nonumber\\
& = \sum_i (g^{(i)}_{WZV})^2 
\left[ 3(M_i^{\pm})^2 - \frac{( M_Z^2 -
M_W^2)^2}{(M_i^{\pm})^2} \right].&
%\end{array}
\eea
Unitarity is violated at a scale 
\be
\Lambda \approx \frac{3\pi^4}{g^2}\frac{M_W^2}{M_1^{\pm}} \approx 5-10
\;{\rm TeV}.
\ee
The first MVB should appear below 1 TeV, and thus accessible at LHC.
In the approximation that the first state $V_1$ saturates the sum rules,
its partial width is given by 
\be 
\Gamma(V_1^{\pm} \to W^{\pm}Z) \approx \frac{\alpha(M_1^{\pm})^3}
{144 s_W^2 M_W^2}.
\ee
For $M_1^{\pm}=700$ GeV, the width is about 15 GeV.
In SM, there is no resonance in $ W^{\pm}Z$ scattering

Even in Higgsless models, one has to distinguish between cases where
$W_LW_L$ scattering is weak, and where it is strong. Thus, if the
MVB's are light, with mass below 1 TeV, gauge-boson scattering will be
weak. However, if the MVB's are far above 1 TeV in mass, $W_LW_L$
scattering will be strong \cite{chanowitz}. 

We now look at the practical aspects of actual observation of $WW$
scattering at the LHC.

\section{\bm $W W$ scattering at LHC}

\begin{figure}[htb]
\begin{center}
\vskip -1cm
\includegraphics[width=8cm]{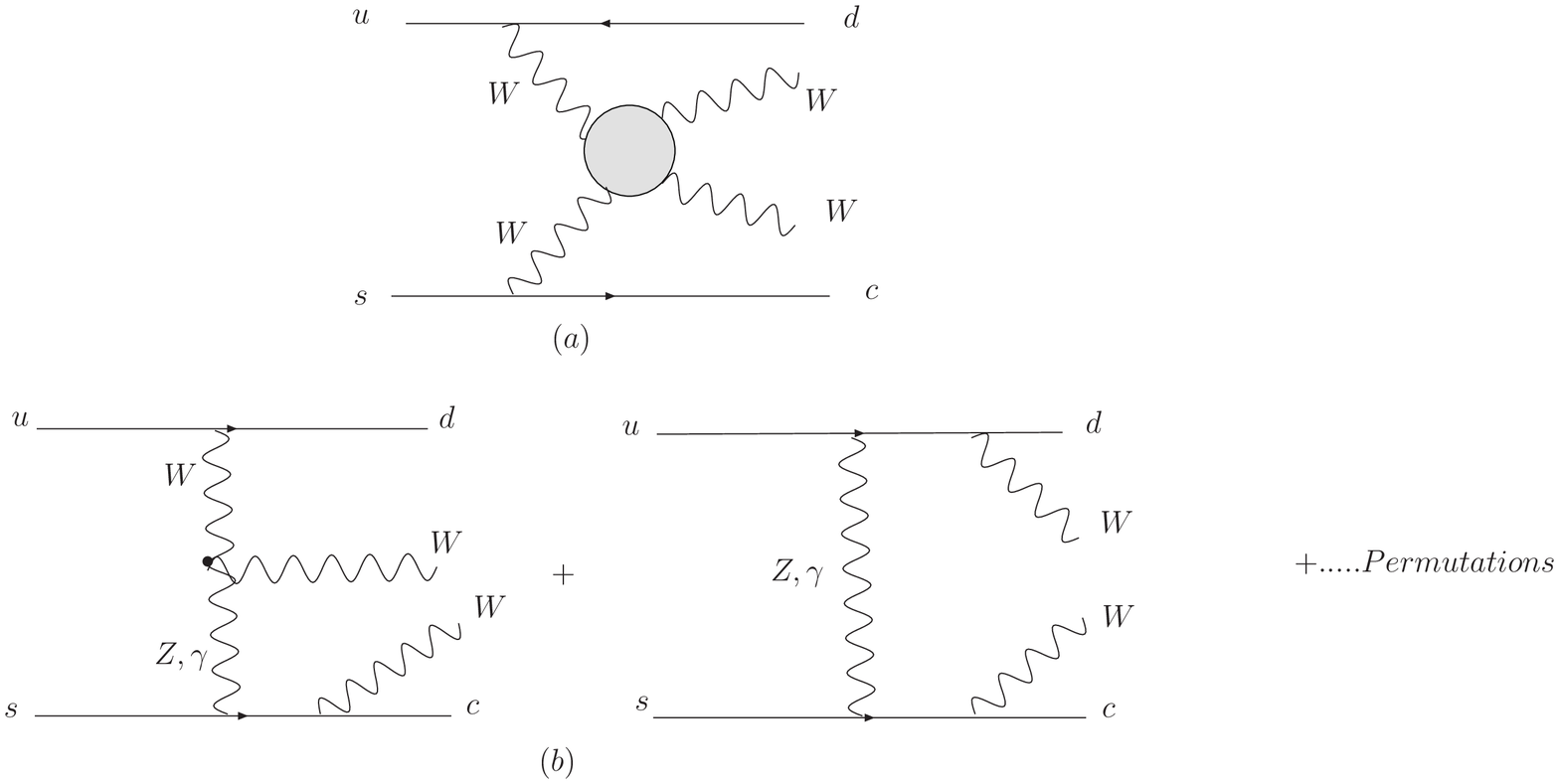}
\vskip -.5cm
\caption{Main diagram topologies for
the process $us \rightarrow cd W^+W^-$. (From \cite{Accomando}).}
\label{uscdww}
\end{center}
\end{figure}

At a hadron collider like the LHC, $WW$ scattering can occur with
virtual $W$'s emitted by the quarks in the hadrons. A $W$ pair in the
final state can be produced either through $WW$ scattering diagrams,
or through $W$ emission from the partons of the initial hadrons. Fig.
\ref{uscdww} shows these two types of contributions. Fig. \ref{uscdww}
(a) represents the genuine $WW$ scattering diagrams, whereas Fig.
\ref{uscdww} (b) shows the ``Bremsstrahlung'' diagrams, which would be a
background in the study of $WW$ scattering. 

Thus, to study $WW$ scattering at LHC, one has to find ways of
separating the genuine scattering contribution from the other
``Bremsstrahlung'' contributions, which is no  mean task. In fact, as we
shall see, the interference between these two classes of diagrams is
large, and huge cancellations can take place. Moreover, the gauge bosons
taking part in the scattering are necessarily virtual (in fact,
space-like). Thus the formalism has to take into account the virtuality
of the gauge bosons. 

A standard approach in the past to processes involving gauge bosons has
been the equivalent vector-boson approximation, an extension of the
equivalent-photon approximation used since early times
\cite{Weizsacker}.

\subsection{\bm Equivalent vector-boson approximation}

The equivalent-photon approximation (Weizs\" acker-Williams approximation)
relates the cross section for a charged particle beam to interact with a
target with a virtual photon exchange
to the cross section for real photon beam to interact with the same
target and produce the final state:
\beq
\sigma = \int dx \sigma_{\gamma}(x) f_{e/\gamma}(x).
\eeq
Here, the photon distribution with momentum fraction $x$ in a 
charged-particle beam of energy $E$ is given by
\beq
f_{e/\gamma}(x) = \frac{\alpha}{2 \pi x}
\ln\left(\frac{E}{m_e}\right) [x + (1-x)^2].
\eeq

This is generalized to what is known as an equivalent (or effective) 
vector-boson
approximation (EVBA) for a process with weak bosons in place of photons
\cite{Dawson,Godbole}. The corresponding distributions of
vector bosons $V$ in a fermion $f$ are given by \cite{Godbole}
\beq
\begin{array}{lcl}
 f_{f/V_{\pm}}(x)& = &\dsp\frac{\alpha}{2 \pi
x}\ln\left(\frac{E}{m_V}\right)
\\ && \!\!\!\!\!\!\times 
\left[( v_f \mp a_f)^2 + (1-x)^2 ( v_f \pm a_f)^2\right]\\
 f_{f/V_L}(x) & = & \dsp \frac{\alpha}{ \pi x}(1-x)
\left[ v_f ^2 + a_f^2\right].
\end{array}
\eeq
Here the suffixes $\pm$ on $V$ denote the helicities $\pm 1$ of $V$ and
$V_L$ is the state with helicity 0.

The use of EVBA entails (a) 
restricting to vector boson scattering
diagrams (b) neglecting diagrams of Bremsstrahlung type
(c) putting on-shell momenta of the vector bosons which take
part in the scattering and 
(d) approximating the total cross section of the process
$f_1f_2\rightarrow f_3f_4V_3V_4$
by the convolution of the vector boson luminosities
$\mathcal {L}^{V_{1}V_{2}}_{s_{1}s_{2}}(x)$ with the on-shell 
cross section:
\be
\begin{array}{ll}
\sigma(f_1f_2\rightarrow f_3f_4V_3V_4)\!\!&= \displaystyle
\int dx\!\!\sum_{V_1,V_2}\sum_{s_{1}s_{2}}
{\mathcal{ L}^{V_{1}V_{2}}_{s_{1}s_{2}}(x)}\\
\!\!&\times 
\sigma^{on}_{s_1s_2}(V_1V_2\rightarrow V_3V_4 ,xs_{qq})
\end{array}
\ee
Here $x=M(V_{1}V_{2})^2/s_{qq}$, while $M(V_{1}V_{2})$ is the vector
boson pair
invariant mass and $s_{qq}$ is the square of the partonic c.m. energy,
and $s_1,s_2$ are the spins of $V_1,V_2$.
Note in the context of (c) above that
the on-shell point
$q^2_{1,2} = M^2_{V_{1,2}}$ is outside the physical 
region $q^2_{1,2}\leq 0$.
Thus the extrapolation involved is more than that involved in the case
of the equivalent-photon approximation.
 
Even if only the longitudinal polarization, expected to be dominant, is kept, 
EVBA {overestimates} the true cross section.
The transverse polarization contribution is found to be comparable to
the longitudinal one \cite{Godbole}.
Improved EVBA \cite{Frederick}, 
going beyond the leading approximation, still
overestimates the cross section \cite{Olness}. 
 Further improvements in EVBA have been attempted  \cite{Kuss}.

\subsection{\bm Backgrounds}

Backgrounds are of two types:

1. Bremsstrahlung processes -- these are processes where the vector
bosons are radiated by quark or anti-quark partons, and 
which do not contribute to $VV$ scattering.

2. Processes which fake a $VV$ final state.

 It is important to understand the first inherent background, and
device cuts which may enhance the signal. 
 However, it may be possible to live with it -- provided $VV$
scattering signal is anyway enhanced because it is strong. In that case,
one simply makes predictions for the combined process of $PP \to VV +
X$.
The second background is crucial to take care of, otherwise we do not
know if we are seeing a $VV$ pair in the final state or not.

Background processes are $q\overline q \to W^+W^-X$, $gg \to
W^+W^-X$,
$t\overline t$ + jet, with top decays giving $ W^+W^-$ pair.
Electroweak-QCD process $W^+$ + jets can mimic the signal when the
invariant mass of the two jets is around $m_W$.
There is a potential background from QCD processes $q\overline q, gg \to t
\overline t X, Wt\overline b$ and ($t\overline t$ +jets), in which a $W$
can come from the decay of $t$ or $\overline t$.
$W$ boson pairs produced from the intrinsic electroweak process
$q\overline q \to q\overline q W^+W^-$ tend to be transversely
polarized.
Coupling to $W^+$ of incoming quark is purely left-handed.
Helicity conservation implies that outgoing quark follows the direction
of incoming quark for longitudinal $W$, and it goes opposite to
direction of incoming quark for transverse (left-handed) $W$.
Hence outgoing quark jet is less forward in background than in signal
event, and tagging of the forward jet can help.

In addition, emission in the central region is favoured in the QCD
background processes, whereas jet production in the central region is
suppressed for $WW$ scattering. Thus, a veto on additional jets in the
central region would be a powerful discriminant between signal an
background. 

For a discussion of the $WW$ scattering in the context of LHC detectors,
see \cite{expt}.

\subsection{\bm Distinguishing the signal}

The feasibility of extracting $WW$ scattering from experiment and
comparison of EVBA with exact results was recently studied by
Accomando and collaborators \cite{Accomando}.
 
It is known that when $W$'s are allowed to be off mass shell,
the amplitude grows faster with energy, as compared to when they are on
shell \cite{Kleiss}.
The problem of bad high-energy behaviour of $WW$  scattering diagrams can be
avoided by the use of the axial gauge \cite{Kunszt}.
%In the axial gauge, Goldstone and gauge fields mix, with the gauge
%propagator given by 
%\beq
%\left(-g_{\mu\nu} + \frac{q_\mu n_\nu + n_\mu q_\nu}{q\cdot n} - \frac
%{n^2}{(q\cdot n)^2}q_\mu q_\nu\right)(q^2-m_W^2)^{-1}.
%\eeq

Accomando et al. \cite{Accomando} have examined
(a) the role of choice of gauge in $WW$ fusion, in particular, the axial
gauge,
(b) the reliability of EVBA,
(c) 
the determination of regions of phase space, in suitable gauge,
which are dominated by the signal (i.e., the $WW$ scattering diagrams).
Their results show that 
 $WW$ scattering diagrams do not constitute the
dominant contribution in any gauge or phase space region.
 Thus, there is no substitute to the complete amplitude for studying 
$WW$ fusion process at LHC.

A similar discussion in the context of the EWCL including resonances, in
contrast to the discussion of Accomando {\it et al.}, is found in
\cite{kilian}, where again the predictions of EVBA are compared with
those obtained from the event generator {\tt WHIZARD}.

We now describe some of the results of 
\cite{Accomando}.

Table \ref{crosx_h} shows the cross section in different gauges for
the contribution of $us$ partons to the full process $pp\to W^+W^-X$, to
only the $WW$ diagrams, and the ratio of these cross sections. The Higgs
boson mass is assumed to be 200
GeV, and a cut on the $WW$ invariant mass $M(WW)$ of 300 GeV is applied.
It is clear that in the $WW$ contribution is largely
cancelled by the ``Bremsstrahlung'' type background contribution. Even in the
axial gauge, in which the $WW$ contribution is the least, it is still a
factor of 2 larger than the actual cross section. 
   %\centerline{ \large $M_H=200$~GeV}
\begin{table}[htb]
   \caption{Contribution to the cross section from $WW$ diagrams and all
diagrams for Higgs mass of
   $M_H$ =200 GeV and $M(WW) >$ 300 GeV. Also shown is the ration of the
$WW$ contribution to the total contribution. (From \cite{Accomando}).}
   \label{crosx_h}
\begin{center}
\begin{tabular}{lcccc}
\hline
%& \multicolumn{1}{|c|}{No-Higgs} \\
& \multicolumn{2}{c}{$\sigma(pb)$}&\\
\cline{2-3}
 Gauge  &   All &  $WW$ & Ratio\\
	&diagrams  &   diagrams & $WW/all$ \\ 
   \hline
   \hline
   Unitary & 8.50 10$^{-3}$   &  6.5 & 765\\
   \hline
   Feynman & 8.50 10$^{-3}$ & 0.221 & 26 \\
   \hline
   Axial & 8.50 10$^{-3}$ & 2.0 10$^{-2}$  & 2.3 \\
   \hline
   \end{tabular}
   \end{center}
   \end{table}

$WW$ invariant-mass
distributions are also obtained in \cite{Accomando}. Again, the result including all 
diagrams does not give a true representation of the
$WW$ contribution alone. Accomando {\it et al.} have also made a 
comparison of the $WW$ invariant mass distribution for the
process $us \rightarrow dc W^+W^-$ in an improved EVBA with the exact complete
result for the two cases of a very heavy Higgs and a Higgs of mass 250
GeV. EVBA exceeds the exact result except at the Higgs resonance.
Ref. \cite{Accomando} also investigated
the total cross section in EVBA and exact
computation and their ratio for different cuts on the $W$ scattering
angle.
The angular cuts serve to decrease the discrepancy between EVBA and
exact computation.

\begin{figure}[htb]\label{largeinvmass}
\begin{center}
\vskip -.5cm
\mbox{\includegraphics[width=6.5cm]{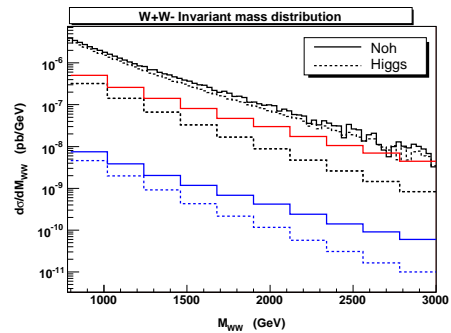}}
\end{center}
\vskip -1cm
\caption{
The $WW$ invariant mass distribution in
$PP\rightarrow us \rightarrow cdW^+W^-$ 
for no Higgs (solid curves) and for $ M_h$=200 GeV
(dashed curve). The two intermediate (red)
curves are obtained imposing cuts shown below.
The two lowest (blue) curves refer to the process
$PP\rightarrow us \rightarrow cd\mu^-\overline{\nu}_\mu e^+\nu_e$ 
with further acceptance cuts: $E_l >
20$
GeV,  $p_{T}^\ell > 10$ GeV , $\vert \eta_l \vert <$ 3.
}
\end{figure}

The $WW$ invariant mass distribution in the process $PP\rightarrow us
\rightarrow cdW^+W^-$ is shown in Fig. \ref{largeinvmass}, imposing the
cuts 
  $E{\rm (quarks)}>20$~GeV, $\pt$(quarks,$W)>10$~GeV,
          $2<\vert\eta{\rm (quark)}\vert<6.5$,
        $\vert\eta {\rm (W)}\vert<3$.
It is seen that for
sufficiently large $WW$ invariant mass it seems feasible to distinguish
between light Higgs and heavy Higgs scenarios. It is reasonable to
anticipate that this is the kinematic region where it may be possible to
test non-standard scenarios of symmetry breaking.

While the study of Accomando {\it et al.} is revealing, it may be
argued that the alternative to SM with a light Higgs boson that they
consider is a somewhat {\it ad hoc} and unphysical $m_H \to \infty$
limit of SM. 
In that context, the study carried out by \' Eboli, Gonzalez-Garcia and 
Mizukoshi \cite{eboli2}
in the framework of EWCL may be considered more systematic. 
They impose jet acceptance cuts of $p_T^j > 20$~GeV, $\vert \eta_j \vert
< 4.9$, and demand a rapidity gap between jets through $\vert \eta_{j1}
- \eta_{j2}\vert > 3.8$, $\eta_{j1}\cdot \eta_{j2} <0$. The lepton
acceptance and isolation cuts used are  $\vert \eta_\ell \vert
< 2.5$, $\eta^j_{\rm min} < \eta_\ell < \eta^j_{\rm max}$, with
$\eta^j_{\rm min(max)}$ as the minimum (maximum) rapidity of the tagging
jets, $\Delta R_{\ell j} \geq 0.4$, $\Delta R_{\ell \ell} \geq 0.4$, 
and $p_T^\ell \geq p_T^{\rm min}$, where $p_T^{\rm min} = 100(30)$~GeV
for opposite(equal) charge leptons. Since signal events contain
neutrinos, they also require a missing transverse momentum $p_T^{\rm
missing} \geq 30$~GeV.
In addition, to suppress the SM background as well as background from
$t\bar t$ production, they impose additional cuts on the invariant
masses of the pair of tagging jets, and of the $W$ pair.
We refer to
the original paper for the details. However, here we limit ourselves to
pointing out the sensitivity they obtain, {\it viz.}, eq.
(\ref{LHCsens}).

The results on the sensitivity obtained in ref. \cite{eboli2} were made
use of in \cite{fabb} to discuss the feasibility of observation of two
scenarios of SSB in the context of EWCL. 
One of these is the heavy Higgs scenario, where the
Higgs mass is assumed to be between 2 and 2.5 TeV. While perturbative
calculations are not reliable for such Higgs masses, it is presumed that
some insight would be obtained into the 
strongly interacting behaviour. A calculation of the EWCL coefficients
in this case yields values which, using eq. (\ref{LHCsens}), are too
small to be observable at LHC. A similar conclusion has been drawn in a
generic technicolor type of model in the large $N$ limit for an $SU(N)$
gauge theory for confinement of techni-fermions.

It seems that the prospects for the observation of non-trivial effects
in $WW$ scattering are limited to vector and scalar resonances, whatever
the dynamics that produce the resonances. A concrete study in the
context of EWCL with extra resonances is \cite{kilian} and in the
context of Higgsless models can be found in \cite{Malhotra}.

\section{Discussion}

In the absence of a light Higgs, $WW$ interactions become strong at TeV scales
leading to violation of perturbative unitarity. 
Study of $WW$ scattering can give information of the electroweak
symmetry breaking sector and discriminate between models.

There are a number of possible scenarios. In the standard model with a
light Higgs boson ($m_H$ \raisebox{-4pt}{$\stackrel{<}{\sim}$} $1$~TeV), $WW$
scattering is well-behaved at large energies. On the other hand, if the
SM Higgs boson is heavier than about 1 TeV, $WW$ interactions become
strong. For extensions of SM, there are various possibilities. One
possibility is that there are no elementary scalars, and SSB is
dynamical in origin. In that case, $WW$ scattering would show new
features which restore unitarity, possibly resonances. 
In theories with extra dimensions like Higgsless models, 
the violation of unitarity is delayed
to higher energies because of the cancellation of the leading
high-energy term by the exchange of Kaluza-Klein excitations of gauge
bosons, which would be seen as resonances. A further
possibility is that there is a light Higgs boson, which postpones
unitarity violation, but there is new physics beyond. The new physics
in its simplest form could be extra Higgs, and $WW$ scattering could be
weak or strong depending on the masses of these Higgs bosons. A more
sophisticated possibility is that the light Higgs is a pseudo-Goldstone
boson of some higher symmetry, and the $WW$ scattering above the Higgs
resonance would show interesting features. Thus $WW$ scattering needs to
be studied even if a light Higgs boson is found. 

New physics could be modelled by means of an effective theory valid at
low energies, {\it viz.}, EWCL, whose lowest dimensional operators
would be fixed by the low-energy theorems, but operators with higher
dimensions would have coefficients fixed either from experiment, or
on the basis of a detailed theory which describes the ultra-violet
sector completely. Such a
formalism has to be combined with a suitable method of unitarization.

Two of the coefficients of higher dimensions are constrained by precision
experiments. Two others would be constrained by the four-point couplings
of the gauge bosons. Studies of popular scenarios of SSB beyond SM show 
that a determination of these latter couplings may pose a
challenge to LHC.

In general there are large
cancellations between the scattering and Bremsstrahlung diagrams.
Hence extraction of $WW$ scattering contribution from the process $PP\to
W^+W^- X$ needs considerable effort.
EVBA overestimates the magnitude in most kinematic distributions.
Appropriate cuts to reduce background are most essential. 
It is possible to extract
information on $WW$ scattering from hadronic experiments by
concentrating on the large-invariant mass region.

\noindent {\bf Acknowledgements} I thank Andreas Nyffeler 
for comments and discussions.
I also thank Namit Mahajan for
discussions and pointing out ref. \cite{oller}.

\end{document}